\documentclass
[superscriptaddress,secnumarabic,amssymb,amsmath,nobibnotes,aps,prd,showkeys,showpacs,nofootinbib,nopacsnumber,onecolumn,12pt]{revtex4}%
\usepackage{graphicx}
\usepackage{bm}
\usepackage{amsmath}
\usepackage{amsfonts}
\usepackage{amssymb}%
\providecommand{\U}[1]{\protect\rule{.1in}{.1in}}

\newcommand{\be}{\begin{equation}}
\newcommand{\ee}{\end{equation}}

\newcommand{\mincir}{\raise
-3.truept\hbox{\rlap{\hbox{$\sim$}}\raise4.truept\hbox{$<$}\ }}
\newcommand{\magcir}{\raise
-3.truept\hbox{\rlap{\hbox{$\sim$}}\raise4.truept\hbox{$>$}\ }}

\begin{document}
\title{Complex Scalar fields in Scalar-Tensor and Scalar-Torsion theories}
\author{Andronikos Paliathanasis}
\email{anpaliat@phys.uoa.gr}
\affiliation{Institute of Systems Science, Durban University of Technology, PO Box 1334,
Durban 4000, South Africa}
\affiliation{Instituto de Ciencias F\'{\i}sicas y Matem\'{a}ticas, Universidad Austral de
Chile, Valdivia 5090000, Chile}
\affiliation{Mathematical Physics and Computational Statistics Research Laboratory,
Department of Environment, Ionian University, Zakinthos 29100, Greece}

\begin{abstract}
We investigate the cosmological dynamics in a spatially flat
Friedmann--Lema\^{\i}tre--Robertson--Walker geometry in scalar-tensor and
scalar-torsion theories where the nonminimally coupled scalar field is a
complex field. We derive the cosmological field equations and we make use of
dimensionless variables in order to determine the stationary points and
determine their stability properties. The physical properties of the
stationary points are discussed while we find that the two-different theories,
scalar-tensor and scalar-torsion theories, share many common features in terms
of the evolution of the physical variables in the background space.

\end{abstract}
\date{\today}
\maketitle

\section{Introduction}

\label{sec1}

Scalar fields in gravitational theory is a simple mechanism to introduce new
degrees of freedom which can play an important role in the description of
observable cosmological phenomena \cite{ratra}. The early acceleration phase
of the universe is attributed to a scalar field known as inflaton
\cite{in0,in1,in2,in3}. In addition, scalar fields have been used to describe
also the late-time acceleration phase of the universe attributed to dark
energy, or other matter components such as dark matter, see for instance
\cite{sf1,sf2,sf3,sf4,sf5,sf6,sf7,sf8,sf9,sf10} and references therein.

Multi-scalar fields have been widely studied in the literature. Some
well-known two-scalar field models are the quintom \cite{qq1} or the Chiral
model which leads to hyperbolic inflation \cite{hy4}, while other proposed
multi-scalar fields theories can be found for instance in
\cite{cc1,mm2,mm3,mm4,mm5,mm6} and references therein. A simple mechanism to
introduce a multi-scalar field theory is to consider the existence of a
complex scalar field, the real and imaginary parts of which give the
equivalent of a two scalar-field theory \cite{com1,com2}.

An inflationary model with a complex scalar field was proposed in \cite{com3}.
Specifically it was found that, when inflation occurs, the imaginary component
of the complex scalar field does not contribute in the cosmological fluid,
that is, the phase of the complex scalar field is constant. The cosmological
perturbations with a complex scalar field were investigated in \cite{com4}.
Furthermore, in \cite{com5} the authors used cosmological observations to
reconstruct the quintessence potential for a complex scalar field. A
nonminimally coupled scalar field cosmological model has been studied in
\cite{com6}, while for some recent studies of complex scalar field
cosmological models we refer the reader to
\cite{com7,com8,com9,com10,com11,com12,com13,com14}.

In this study we consider the scalar-tensor and the scalar-torsion theories
with a complex scalar field \cite{sc1}. In scalar-tensor theory, the scalar
field is minimally coupled to gravity. The scalar field interacts with the
gravitational Action Integral of Einstein's General Relativity, that is, the
Ricci-scalar of the Levi-Civita connection, the scalar-tensor theory satisfies
the Machian Principle \cite{Brans} and the theory is defined in the so-called
Jordan frame \cite{Jord}. The Brans-Dicke Lagrangian \cite{Brans} is the most
common scalar-tensor theory. On the other hand, the scalar-torsion theory is
the equivalent of scalar-tensor model in teleparallelism. In the latter, the
fundamental geometric invariant is the torsion scalar determined by the
curvatureless Weitzenb\"{o}ck connection \cite{Hayashi79,Tsamp}. There are
many important results in the literature on the cosmological studies of the
scalar-tensor and scalar-torsion theories, for an extended discussion we refer
the reader to \cite{omegaBDGR,st1,st2,st3,st4,st5,st6,st7,st8,te2,te3,te4} and
references therein.

The purpose of this study is to investigate the effects of a complex scalar
field in the evolution of the cosmological dynamics for the background space
for these two different gravitational theories and to compare the results.
Such analysis provides us with important results in order to understand the
differences between the use of the Ricci-scalar and of the torsion scalar in
the background geometry. We use dimensionless variables to perform a detailed
analysis of the dynamical systems which describe the evolution of the physical
variables. Such an approach has been widely studied before with many
interesting results about the viability of proposed gravitational theories,
see for instance \cite{dyn1,dyn2,dyn3,dyn4,dyn6,dyn5}. The plan of the paper
is as follows.

In Section \ref{sec2} we consider a spatially flat
Friedmann--Lema\^{\i}tre--Robertson--Walker geometry in scalar-tensor
gravitational theory with complex scalar field. We derive the field equations
and we write the point-like Lagrangian. Moreover, we determined the stationary
points for the field equations and we investigate their stability properties.
In Section \ref{sec3} we perform a similar analysis but now in the context of
scalar-tensor theory. Finally, in Section \ref{conc} we summarize our results
and compare the physical results provided by the two theories and draw our conclusions.

\section{Scalar-Tensor Cosmology}

\label{sec2}

Consider the complex scalar field $\psi$ in the case of scalar-tensor theory
\cite{sc1}, for which the gravitational Action Integral is defined as
\begin{equation}
S_{Tensor}=\int d^{4}x\sqrt{-g}\left[  F\left(  \left\vert \psi\right\vert
\right)  R+\frac{1}{2}g^{\mu\nu}\psi_{,\mu}\psi_{,\nu}^{\ast}-V\left(
\left\vert \psi\right\vert \right)  \right]  , \label{ai.01}%
\end{equation}
where $R$ is the Ricci scalar related to the Levi-Civita connection for the
metric tensor $g_{\mu\nu}$; $\left\vert \psi\right\vert $ is the norm of the
complex field $\psi$, that is, $\left\vert \psi\right\vert ^{2}=\psi\psi
^{\ast}$, $F(\left\vert \psi\right\vert )$ is the coupling function between
the gravitational and the scalar field $\psi$ and $V\left(  \left\vert
\psi\right\vert \right)  $ is the potential function which drives the
dynamics. The Action Integral (\ref{ai.01}) admits the $U\left(  1\right)  $ symmetry.

The Brans-Dicke theory with a complex field is recovered for~$F(\left\vert
\psi\right\vert )=F_{0}\left\vert \psi\right\vert ^{2}$, that is, the Action
Integral (\ref{ai.01}) is \cite{sc1}
\begin{equation}
S_{BD}=\int d^{4}x\sqrt{-g}\left[  F_{0}\left\vert \psi\right\vert ^{2}%
R+\frac{1}{2}g^{\mu\nu}\psi_{,\mu}\psi_{,\nu}^{\ast}-V\left(  \left\vert
\psi\right\vert \right)  \right]  . \label{ai.02}%
\end{equation}

In the case of a spatially flat FLRW universe with line element,%
\begin{equation}
ds^{2}=-N^{2}\left(  t\right)  dt^{2}+a^{2}\left(  t\right)  \left(
dx^{2}+dy^{2}+dz^{2}\right)  ~, \label{ai.03}%
\end{equation}
we derive for the Ricciscalar
\begin{equation}
R=6\left(  \frac{1}{N}\dot{H}+12H^{2}\right)  ~,~ \label{ai.04}%
\end{equation}
where $H=\frac{1}{N}\frac{\dot{a}}{a}$, $\dot{a}=\frac{da}{dt}$, is the Hubble function.

We substitute (\ref{ai.04}) into (\ref{ai.01}) which by integration by parts
gives the point-like Lagrangian%
\begin{equation}
L_{Tensor}\left(  N,a,\dot{a},\psi,\dot{\psi}\right)  =\frac{1}{N}\left(
6F\left(  \left\vert \psi\right\vert \right)  a\dot{a}^{2}+6\dot{F}\left(
\left\vert \psi\right\vert \right)  a^{2}\dot{a}+\frac{1}{2}a^{3}\left(
\dot{\psi}\dot{\psi}^{\ast}\right)  \right)  -a^{3}NV\left(  \left\vert
\psi\right\vert \right)  , \label{ai.05}%
\end{equation}
or in the case of Brans-Dicke%
\begin{equation}
L_{BD}\left(  N,a,\dot{a},\psi,\dot{\psi}\right)  =\frac{1}{N}\left(
6F_{0}\left\vert \psi\right\vert ^{2}a\dot{a}^{2}+6F_{0}\left(  \left\vert
\psi\right\vert ^{2}\right)  ^{\cdot}a^{2}\dot{a}+\frac{1}{2}a^{3}\left(
\dot{\psi}\dot{\psi}^{\ast}\right)  \right)  -a^{3}NV\left(  \left\vert
\psi\right\vert \right)  . \label{ai.06}%
\end{equation}

We focus now in the Brans-Dicke theory in which the field equations are
described by the point-like Lagrangian (\ref{ai.06}). Moreover, the complex
scalar field is written with the use of the polar form, $\psi\left(  t\right)
=\phi\left(  t\right)  e^{i\theta\left(  t\right)  }$, such that the
Lagrangian (\ref{ai.06}) becomes%
\begin{equation}
L_{BD}\left(  N,a,\dot{a},\phi,\dot{\phi},\theta,\dot{\theta}\right)
=\frac{1}{N}\left(  6F_{0}\phi^{2}a\dot{a}^{2}+12F\phi a^{2}\dot{a}\dot{\phi
}+\frac{1}{2}a^{3}\left(  \dot{\phi}^{2}+\phi^{2}\dot{\theta}^{2}\right)
\right)  -a^{3}NV\left(  \phi\right)  . \label{ai.07}%
\end{equation}

It is obvious that the Lagrangian function (\ref{ai.07}) describes a
multi-scalar field cosmological model, where $\phi$ is the Brans-Dicke field
and $\theta$ is a second-scalar field minimally coupled to gravity but coupled
to the Brans-Dicke field $\phi$. The $U\left(  1\right)  $ for the point-like
Lagrangian (\ref{ai.07}) provides the invariant transformation $\theta
=\theta+\varepsilon$, and the conservation law $I_{0}=\frac{1}{N}a^{3}\phi
^{2}\dot{\theta}$.

Variation with respect to the dynamical variables $\left\{  N,a,\phi
,\theta\right\}  $ of the Lagrangian (\ref{ai.07}) provides the cosmological
field equations which are%
\begin{equation}
0=6F_{0}\phi^{2}H^{2}+12F\phi H\dot{\phi}+\frac{1}{2}\left(  \dot{\phi}%
^{2}+\phi^{2}\dot{\theta}^{2}\right)  +V\left(  \phi\right)  ~, \label{ai.08}%
\end{equation}%
\begin{equation}
0=2F_{0}\phi^{2}\left(  2\dot{H}+3H^{2}\right)  +8F_{0}H\phi\dot{\phi}%
-\frac{1}{2}\dot{\phi}^{2}+4F_{0}\dot{\phi}^{2}-\frac{1}{2}\dot{\phi}^{2}%
\dot{\theta}^{2}+4F_{0}\phi\ddot{\phi}+V\left(  \phi\right)  ~, \label{ai.09}%
\end{equation}%
\begin{equation}
0=\ddot{\phi}+\phi\left(  12F_{0}\dot{H}-\dot{\theta}^{2}\right)  +3\dot{\phi
}\dot{H}+12F_{0}\phi H^{2}+V_{,\phi}~ \label{ai.10}%
\end{equation}
and%
\begin{equation}
0=\phi\ddot{\theta}+\left(  2\dot{\phi}+3H\phi\right)  \dot{\theta}~\text{or
}I_{0}=a^{3}\phi^{2}\dot{\theta}\text{,} \label{ai.11}%
\end{equation}
where without loss of generality we have assumed $N\left(  t\right)  =1$.

\subsection{Cosmological dynamics}

In order to reconstruct the cosmological history provided by this specific
complex scalar-tensor theory we make use of dimensionless variables in the
context of $H$-normalization and we investigate the dynamical evolution of the
field equations (\ref{ai.08})-(\ref{ai.11}) by determining the stationary
points and their stability properties.

We consider the new dimensionless dependent variables \cite{dyn1}%
\begin{equation}
\dot{\phi}=2\sqrt{3}H\phi x~,~V\left(  \phi\right)  =6H^{2}\phi^{2}%
y~,~\dot{\psi}=2\sqrt{3}Hz~,~V_{,\phi}=\lambda\frac{V\left(  \phi\right)
}{\phi}. \label{ai.12}%
\end{equation}
Hence, for the new independent variable, $\tau=\ln a$, the field equations
(\ref{ai.08})-(\ref{ai.11}) are written as the following dynamical system%
\begin{align}
2F_{0}\left(  12F_{0}-1\right)  \frac{dx}{d\tau}  &  =2\sqrt{3}\left(
7-48F_{0}\right)  x^{2}+\left(  3-24F_{0}\right)  x^{3}\label{ai.13}\\
&  +2\sqrt{3}F_{0}\left(  F_{0}+z^{2}+\left(  \lambda-3\right)  y\right)
+3x\left(  F_{0}+\left(  4\lambda F_{0}-1\right)  y+\left(  1-8F_{0}\right)
z^{2}\right)  ~,\nonumber
\end{align}%
\begin{align}
F_{0}\left(  12F_{0}-1\right)  \frac{1}{y}\frac{dy}{d\tau}  &  =2\sqrt{3}%
F_{0}x\left(  4-\lambda+12F_{0}\left(  \lambda-2\right)  \right)  +\left(
3-24F_{0}\right)  x^{2}\label{ai.14}\\
&  +3\left(  F_{0}\left(  16F_{0}-1\right)  +\left(  4\lambda F_{0}-1\right)
y+\left(  1-8F_{0}\right)  z^{2}\right)  ~,\nonumber
\end{align}%
\begin{align}
-2F_{0}\left(  12F_{0}-1\right)  \frac{1}{3z}\frac{dz}{d\tau}  &  =4\sqrt
{3}F_{0}\left(  8F_{0}-1\right)  x+\left(  8F_{0}-1\right)  x^{2}%
\label{ai.15}\\
&  +\left(  1-4\lambda F_{0}\right)  y+\left(  8F_{0}-1\right)  \left(
F_{0}+z^{2}\right)  ~,\nonumber
\end{align}%
\begin{equation}
\frac{d\lambda}{d\tau}=2\sqrt{3}\lambda x\left(  1-\lambda+\Gamma\left(
\lambda\right)  \right)  ~,~\Gamma\left(  \lambda\right)  =\phi\frac
{V_{,\phi\phi}}{V_{,\phi}} \label{ai.16}%
\end{equation}
with algebraic constraint%
\begin{equation}
x^{2}+z^{2}+y+F_{0}\left(  1+4\sqrt{3}x\right)  =0~. \label{ai.17}%
\end{equation}

For the scalar field potential we assume $V\left(  \phi\right)  =V_{0}%
\phi^{\lambda_{0}}$; such that $\lambda=\lambda_{0}$ is always a constant and
the dimension of the dynamical system is two, after we apply the constraint
equation (\ref{ai.17}). From (\ref{ai.17}) we substitute the parameter $y$ and
we end up with a two-dimensional system on the space of variables $\left\{
x,z\right\}  $. Moreover, we observe that the dynamical system remains
invariant under the change of variable $z\rightarrow-z$. Thus, without loss of
generality, we can restrict our analysis to the region with $z\geq0$.
Furthermore, for completeness of our analysis the scalar field potential can
be positive or negative such that $y\in%
\mathbb{R}
$.

The stationary points of the reduced dynamical system $A=\left(  x\left(
A\right)  ,z\left(  A\right)  \right)  $ are as follows%
\[
A_{1}=\left(  -\frac{\sqrt{3}F_{0}\left(  \lambda-4\right)  }{6F_{0}\left(
\lambda+2\right)  -3},0\right)  ~
\]
and%
\[
A_{2}=\left(  x_{2},\sqrt{-F_{0}-x_{2}\left(  4\sqrt{3}F_{0}+x_{2}\right)
}\right)  \text{\thinspace}.
\]

Each stationary point describes an asymptotic solution for the field equations
in which the effective fluid has the equation of state parameter%
\begin{equation}
w_{eff}\left(  x,y,z\right)  =\frac{3\sqrt{3}F_{0}x+\left(  3-34F_{0}\right)
x^{2}+3\left(  4F_{0}^{2}+\left(  4F_{0}\lambda-1\right)  y+\left(
1-8F_{0}\right)  z^{2}\right)  }{3F_{0}\left(  12F_{0}-1\right)  }.
\label{ai.18}%
\end{equation}
We proceed with the discussion of the physical properties and the stability
properties for the admitted stationary points.

\begin{figure}[ptb]
\centering\includegraphics[width=1\textwidth]{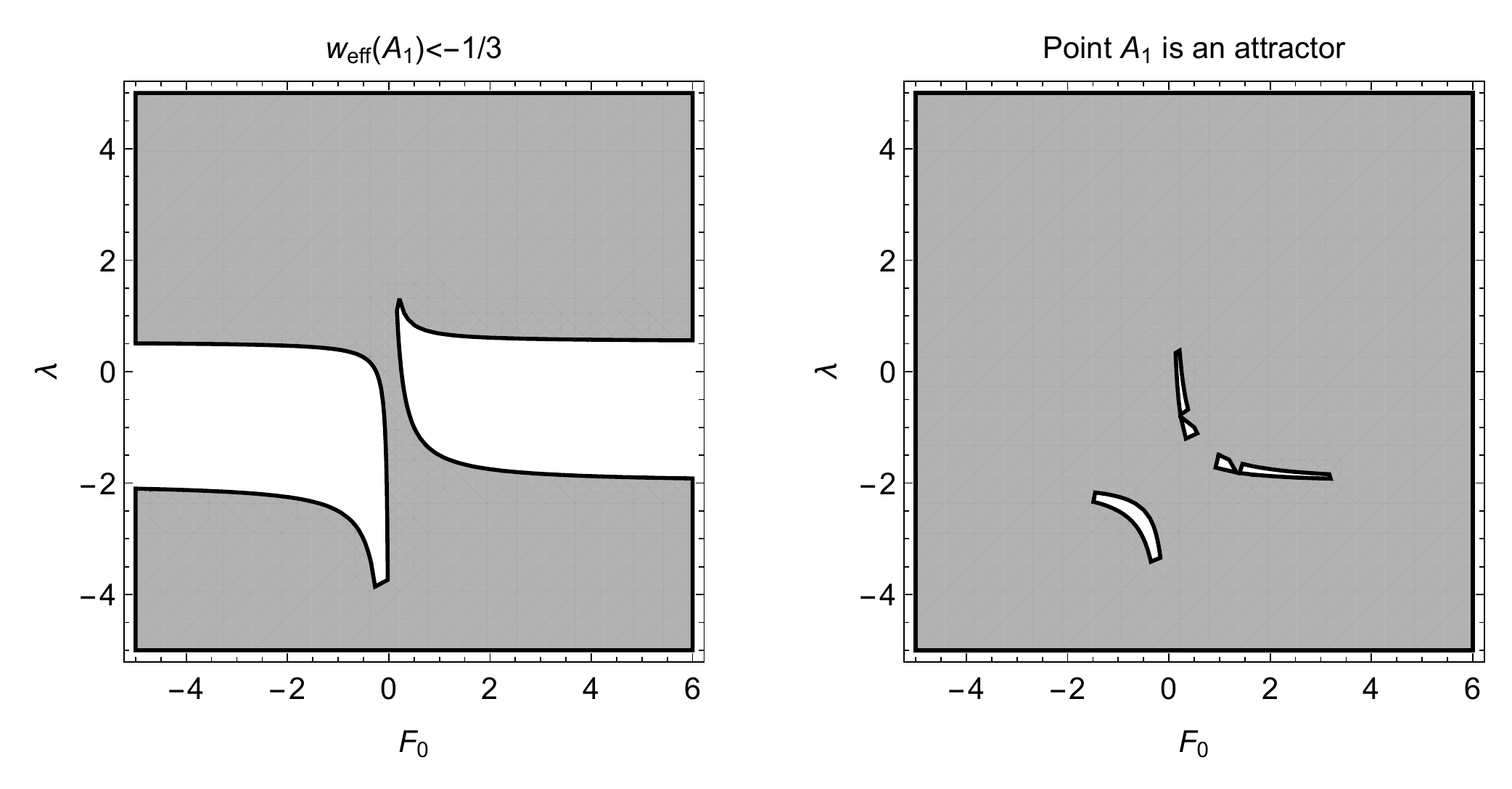}\caption{Region in the
space of the free variables $\left\{  F_{0},\lambda\right\}  $, where
$w_{eff}\left(  A_{1}\right)  <-\frac{1}{3}$ (Left Fig.) and where $A_{1}$ is
an attractor (Right Fig.). It is easy to observe that in this specific region
when $A_{1}$ describes an accelerated universe is also an attractor for the
dynamical system. }%
\label{ff01}%
\end{figure}

The stationary point $A_{1}$ describes a universe dominated only by the scalar
field $\phi$ and its potential function $V\left(  \phi\right)  $, because
$y\left(  A_{1}\right)  =\frac{F_{0}\left(  12F_{0}-1\right)  \left(
3+F_{0}\left(  \lambda-10\right)  \left(  2+\lambda\right)  \right)
}{3\left(  1-2F_{0}\left(  2+\lambda\right)  \right)  ^{2}}$. The effective
equation of state parameter for the asymptotic solution is $w_{eff}\left(
A_{1}\right)  =\frac{3+2F_{0}\left(  2+\lambda\left(  \lambda-9\right)
\right)  }{6F_{0}\left(  2+\lambda\right)  -3}$, from which it follows that
$A_{1}$ describes an accelerated universe as is given in Fig. \ref{ff01}.
Moreover, the eigenvalues of the linearized system around the stationary point
are derived, $e_{1}\left(  A_{1}\right)  =\frac{3+F_{0}\left(  \lambda
-10\right)  \left(  2+\lambda\right)  }{2F_{0}\left(  2+\lambda\right)  -1}%
~$,$~e_{2}\left(  A_{1}\right)  =\frac{3+F_{0}\left(  \lambda-10\right)
\left(  2+\lambda\right)  }{2F_{0}\left(  2+\lambda\right)  -1}$.In Fig.
\ref{ff01} we present the region plot in the two-dimensional space of the free
variables $\left\{  F_{0},\lambda\right\}  $, where the eigenvalues have
negative real parts, that is, point $A_{1}$ is an attractor.

\begin{figure}[ptb]
\centering\includegraphics[width=1\textwidth]{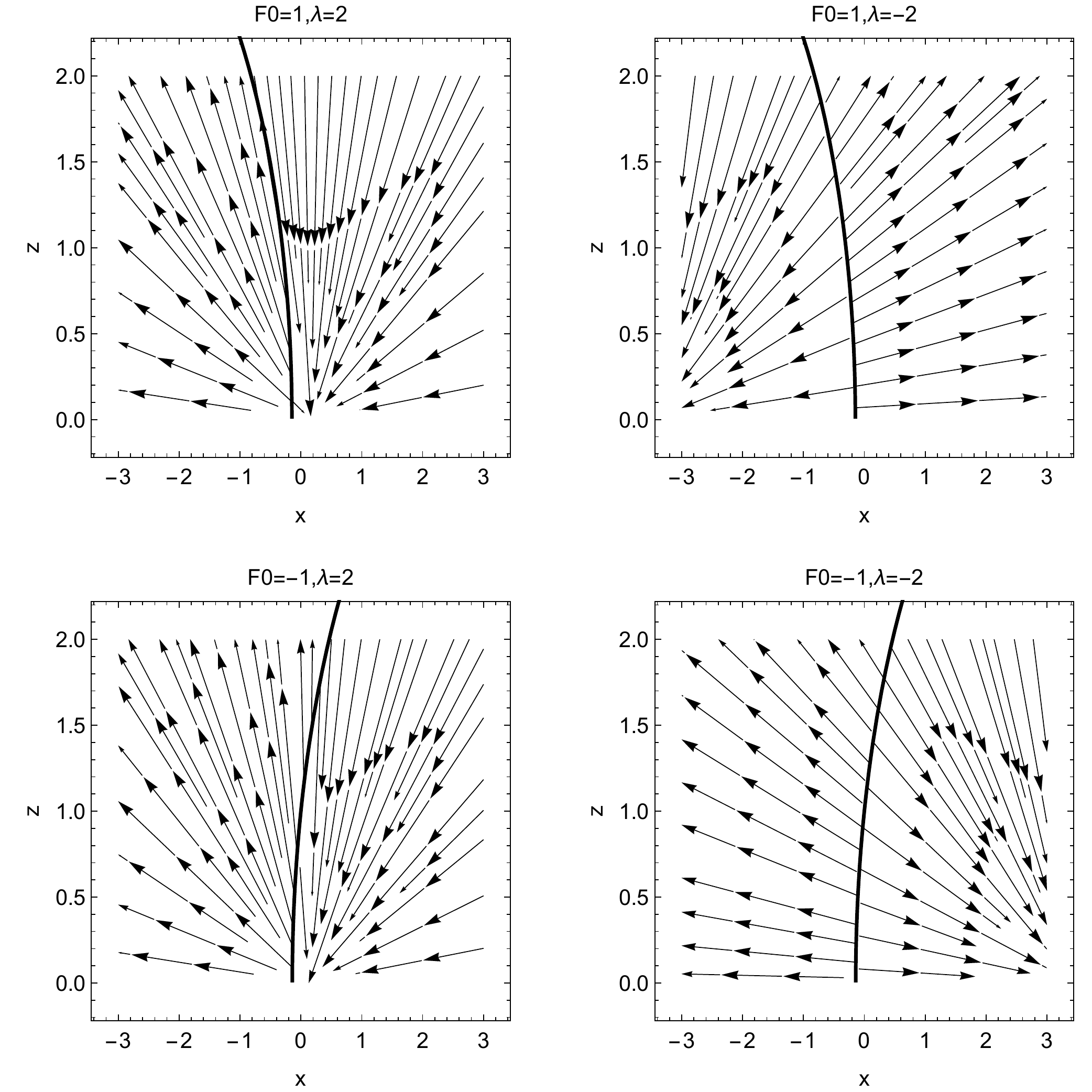}\caption{Phase-space
portrait for the scalar-tensor theory on the two-dimensional plane $\left(
x,z\right)  $ for different values of the free parameters $F_{0}$ and
$\lambda$. From the figure it is clear that the surface of points described by
$A_{2}$ are always saddle points or source points. With solid line we present
the family of points $A_{2}$ for $e_{1}\left(  A_{2}\right)  <0$.}%
\label{f12}%
\end{figure}

Points $A_{2}$ describe a family of stationary points which are real and
physical accepted when $-F_{0}-x_{2}\left(  4\sqrt{3}F_{0}+x_{2}\right)  >0$.
For the asymptotic solution we calculate $w_{eff}\left(  A_{2}\right)
=1+\frac{8}{\sqrt{3}}x_{2}$, which means that the family of points $A_{2}$
describe accelerated universes for $x_{2}<-\frac{1}{2\sqrt{3}}$. Furthermore,
the eigenvalues are calculated $e_{1}\left(  A_{2}\right)  =6+2\sqrt{3}\left(
2+\lambda\right)  x_{2}$ and $e_{2}\left(  A_{2}\right)  =0$. Then for
$6+2\sqrt{3}\left(  2+\lambda\right)  x_{2}<0$, where $e_{1}\left(
A_{2}\right)  <0$, the \ Center Manifold Theorem (CMT) can be applied in order
to investigate for a possible stable submanifold and to infer about the
stability. However, such an analysis does not contribute to the physical
discussion of the present theory and we select to work numerically. In Fig.
\ref{f12} we present the phase space portrait in the two-dimensional place
$\left(  x,z\right)  $ from which we observe that the stationary points are
always saddle points for $e_{1}\left(  A_{2}\right)  <0$.

From the above analysis it is clear that various major eras of the
cosmological evolution can be described by the stationary points, $A_{1}$ and
$A_{2}$. For instance for $F_{0}=-\frac{3}{2}\left(  2-9\lambda+\lambda
^{2}\right)  ^{-1}$, $w_{eff}\left(  A_{1}\right)  =0$ which means that
$A_{1}$ describes the matter dominated era, and points $A_{2}$ can describe
the early and late acceleration phases of the universe. That is not the unique
case, since, for $x_{2}=-\frac{\sqrt{3}}{8},$~$w_{eff}\left(  A_{2}\right)
=0,$ and $A_{1}$ can describe the future acceleration phase of the universe.

\subsection{Analysis at Infinity}

Because variables $\left(  x,z\right)  $ are not bounded, they can take values
at all the range of the real numbers, which means that they can take values
at  infinity. Until now we have investigated the stationary points at the
finite regime. Hence in order to search for stationary points at infinity we
consider the Poincar\'e map%
\begin{equation}
x=\frac{X}{\sqrt{1-X^{2}-Z^{2}}}~,~z=\frac{Z}{\sqrt{1-X^{2}-Z^{2}}}%
~,~d\sigma=\sqrt{1-X^{2}-Z^{2}}d\tau. \label{ai.P}%
\end{equation}

Therefore, at infinity, $1-X^{2}-Z^{2}=0$, i.e. $Z=\sqrt{1-X^{2}}$, the
dynamical system is reduced to the single ordinary differential equation%
\begin{equation}
\frac{dX}{d\sigma}=\frac{\sqrt{3}\left(  \lambda-4\right)  }{12F_{0}-1}\left(
X^{2}-1\right)  .
\end{equation}

Consequently, the stationary points at the infinity $B=\left(  X\left(
B\right)  ,Z\left(  B\right)  \right)  $ are
\begin{equation}
B_{1}=\left(  1,0\right)  \text{ and }B_{2}=\left(  -1,0\right)  \text{.}%
\end{equation}

Hence, at infinity only the scalar field $\phi$ contributes to the
cosmological solution and the physical properties of the points are similar to
those of $A_{1}$.

The eigenvalues of the two-dimensional dynamical system at the stationary
points at infinity are%
\begin{equation}
e_{1}\left(  B_{1}\right)  =\frac{\sqrt{3}\left(  \lambda-4\right)  }%
{12F_{0}-1}~,~\operatorname{Re}\left(  e_{2}\left(  B_{1}\right)  \right)  =0
\end{equation}
and%
\begin{equation}
e_{1}\left(  B_{2}\right)  =-\frac{\sqrt{3}\left(  \lambda-4\right)  }%
{12F_{0}-1}~,~\operatorname{Re}\left(  e_{2}\left(  B_{2}\right)  \right)  =0.
\end{equation}

\begin{figure}[ptb]
\centering\includegraphics[width=1\textwidth]{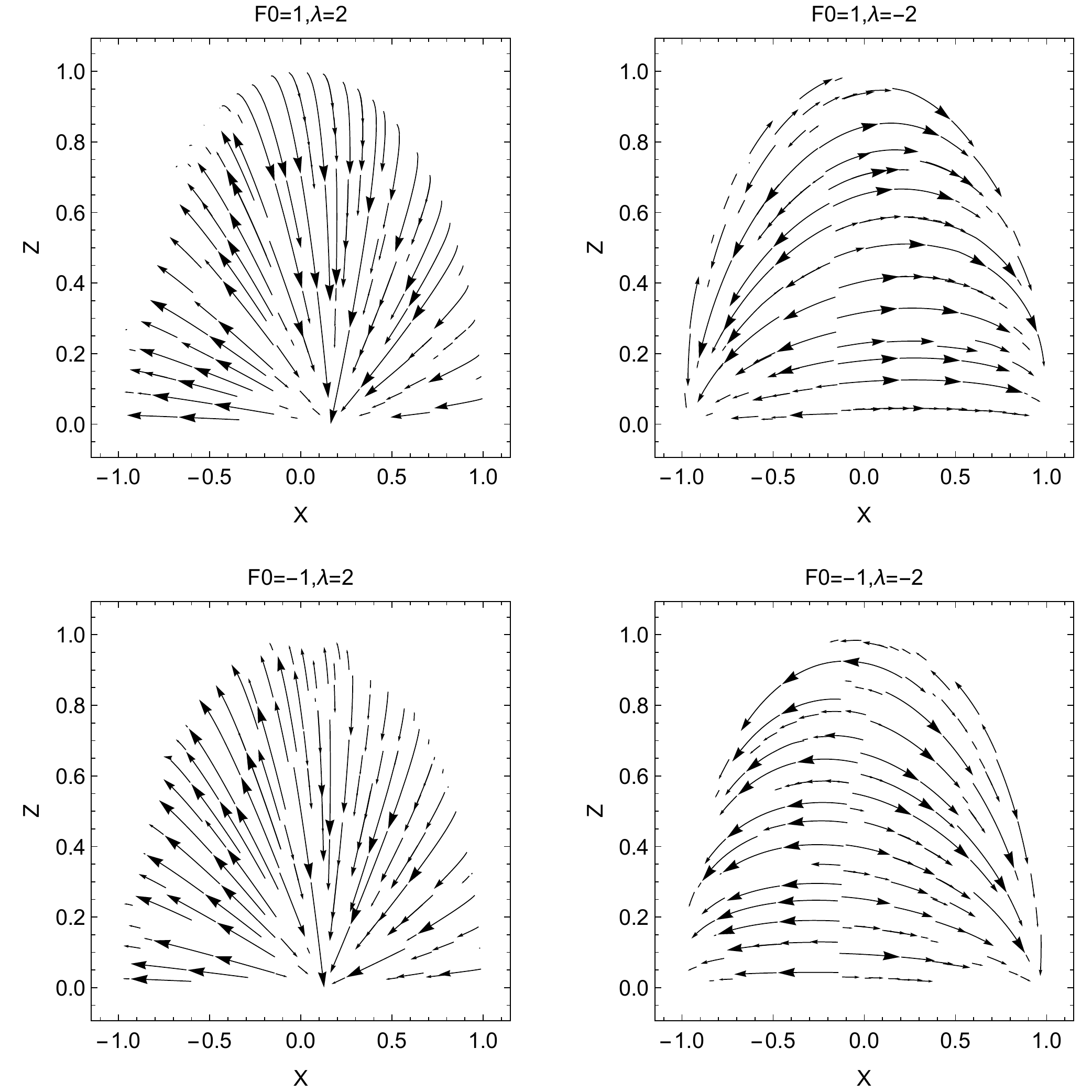}\caption{Phase-space
portrait for the scalar-tensor theory on the two-dimensional plane $\left(
X,Z\right)  $ for $Z\geq0,$ for different values of the free parameters
$F_{0}$ and $\lambda$. From the phase-space portaits we observe that points
$B_{1}$ and $B_{2}$ can be attractors at the infinity regime for the dynamical
system. }%
\label{ff13}%
\end{figure}

From the phase-space portraits of Fig. \ref{ff13} we can easily observe that
stationary point $B_{1}$ is an attractor when $e_{1}\left(  B_{1}\right)  <0$,
while, when $e_{1}\left(  B_{1}\right)  >0,$ point $B_{2}$ is an attractor.

\section{Scalar-Torsion Cosmology}

\label{sec3}

Scalar-torsion theory can be seen as the extension of scalar-tensor theory in
teleparallelism.\ The gravitational Action Integral is \cite{te2}
\begin{equation}
S_{Torsion}=\int d^{4}x\sqrt{-e}\left[  F\left(  \left\vert \psi\right\vert
\right)  T+\frac{1}{2}g^{\mu\nu}\psi_{,\mu}\psi_{,\nu}^{\ast}-V\left(
\left\vert \psi\right\vert \right)  \right]  , \label{ai.20}%
\end{equation}
where $T$ is the torsion scalar of the antisymmetric Weitzenb\"{o}ck
connection, $e=\sqrt{-g}$, $e_{i}$ describes the unholonomic frame, with
$g(e_{i},e_{j})=\mathbf{e}_{i}\cdot\mathbf{e}_{i}=\eta_{ij}$ or in terms of
coordinates, $e_{i}=h_{i}^{\mu}\left(  x^{\kappa}\right)  \partial_{i}$, where
now the Weitzenb\"{o}ck connection is expressed as \cite{revtel}
\begin{equation}
\hat{\Gamma}^{\lambda}{}_{\mu\nu}=h_{a}^{\lambda}\partial_{\mu}h_{\nu}^{a}~
\end{equation}
and $T={S_{\beta}}^{\mu\nu}{T^{\beta}}_{\mu\nu}$, with $T_{\mu\nu}^{\beta}$ to
be the torsion tensor $T_{\mu\nu}^{\beta}=\hat{\Gamma}_{\nu\mu}^{\beta}%
-\hat{\Gamma}_{\mu\nu}^{\beta}$ and
\begin{equation}
{S_{\beta}}^{\mu\nu}=\frac{1}{2}({K^{\mu\nu}}_{\beta}+\delta_{\beta}^{\mu
}{T^{\theta\nu}}_{\theta}-\delta_{\beta}^{\nu}{T^{\theta\mu}}_{\theta}),
\end{equation}
where ${K^{\mu\nu}}_{\beta}$ is the contorsion tensor
\begin{equation}
{K^{\mu\nu}}_{\beta}=-\frac{1}{2}({T^{\mu\nu}}_{\beta}-{T^{\nu\mu}}_{\beta
}-{T_{\beta}}^{\mu\nu}).
\end{equation}

For the spatially flat FLRW spacetime with line element (\ref{ai.03}) we
consider the diagonal frame
\begin{equation}
h_{~\mu}^{i}(t)=\mathrm{diag}(N\left(  t\right)  ,a(t),a(t),a(t)),
\end{equation}
from which we calculate \cite{revtel}
\begin{equation}
T=-6H^{2}.
\end{equation}

By replacing in\ the Action Integral (\ref{ai.20}) and substituting the
complex scalar field as $\psi=\phi e^{i\theta}$, we end up with the point-like
Lagrangian function%
\begin{equation}
L_{Torsion}\left(  N,a,\dot{a},\phi,\dot{\phi},\psi,\dot{\psi}\right)
=-\frac{6}{N}F\left(  \phi\right)  a\dot{a}^{2}+\frac{a^{3}}{2N}\left(
\dot{\phi}^{2}+\phi\dot{\theta}^{2}\right)  -a^{3}NV\left(  \phi\right)  .
\label{ai.21}%
\end{equation}

In an analogue to the Brans-Dicke model we assume now that $F\left(
\phi\right)  =F_{0}\phi^{2}$. This is not a random choice, the Dilaton field,
or the Brans-Dicke field for the specific potential function $V\left(
\phi\right)  =V_{0}\phi^{2}$ admits a discrete symmetry known as the
Gasperini-Veneziano duality transformation. Similarly, the scalar-torsion
theory for $F\left(  \phi\right)  =F_{0}\phi^{2}$ and potential function
$V\left(  \phi\right)  =V_{0}\phi^{2}$ admits a discrete symmetry similar to
the Gasperini-Veneziano duality transformation. Thus, the $F\left(
\phi\right)  =F_{0}\phi^{2}$ can be seen as the teleparallel Brans-Dicke
equivalent model.

Variation with respect to the dynamical variables in the Lagrangian function
(\ref{ai.21}) gives the field equations
\begin{equation}
0=-6F_{0}\phi^{2}H^{2}+\frac{1}{2}\left(  \dot{\phi}^{2}+\phi^{2}\dot{\theta
}^{2}\right)  +V\left(  \phi\right)  ~, \label{ai.22}%
\end{equation}%
\begin{equation}
4F_{0}\phi^{2}\left(  2\dot{H}+3H^{2}\right)  +16F_{0}\phi H\dot{\phi}+\left(
\dot{\phi}^{2}+\phi^{2}\dot{\theta}^{2}\right)  -V\left(  \phi\right)  =0~,
\label{ai.23}%
\end{equation}%
\begin{equation}
\ddot{\phi}-\phi\dot{\theta}^{2}+3H\left(  4F_{0}H\phi+\dot{\phi}\right)
+V_{,\phi}=0~ \label{ai.24}%
\end{equation}
and
\begin{equation}
\phi\ddot{\theta}+\dot{\theta}\left(  3H\phi+2\dot{\phi}\right)  =0~.
\label{ai.25}%
\end{equation}

As in the case of the scalar-tensor theory we proceed with the analysis for
the dynamics of the latter system of differential equations.

\subsection{Cosmological dynamics}

We make use of the dimensionless variables (\ref{ai.12}) and we write the
field equations (\ref{ai.22})-(\ref{ai.25}) in the equivalent dynamical system%
\begin{equation}
F_{0}\frac{dx}{d\tau}=\left(  4\sqrt{3}F_{0}+3x\right)  x^{2}-2\sqrt{3}%
F_{0}\left(  2F_{0}+\lambda y-2z^{2}\right)  -3x\left(  F_{0}+y-z^{2}\right)
~,~
\end{equation}%
\begin{equation}
F_{0}\frac{dy}{d\tau}=y\left(  2\sqrt{3}F_{0}\left(  2+\lambda\right)
x+3x^{2}+3\left(  F_{0}-y+z^{2}\right)  \right)  ~,
\end{equation}%
\begin{equation}
F_{0}\frac{dz}{d\tau}=\frac{3}{2}z\left(  x^{2}-y+z^{2}-F_{0}\right)  ~,
\end{equation}%
\begin{equation}
\frac{d\lambda}{d\tau}=2\sqrt{3}\lambda x\left(  1-\lambda+\Gamma\left(
\lambda\right)  \right)  ~,~~\Gamma\left(  \lambda\right)  =\phi\frac
{V_{,\phi\phi}}{V_{,\phi}}%
\end{equation}
with constraint equation%
\begin{equation}
x^{2}+y+z^{2}-F_{0}=0, \label{ai.30}%
\end{equation}
where now the parameter for the equation of state for the effective fluid is
expressed as follows%
\begin{equation}
w_{eff}\left(  x,y,z\right)  =\frac{8}{\sqrt{3}}x+\frac{1}{F_{0}}\left(
x^{2}-y+z^{2}\right)  \text{.}%
\end{equation}

We observe that the field equations are invariant under the change of variable
$z\rightarrow-z$. Thus without loss of generality we select to work in the
region $z\geq0$. Moreover, from the constraint equation (\ref{ai.30}) the
dimension of the dynamical system is reduced by one, while, for the power-law
potential function $V\left(  \phi\right)  =V_{0}\phi^{\lambda}$, the field
dimension of the dynamical system is two.

The stationary points on the two-dimensional plane $\left\{  x,z\right\}  $,
are of the form $C=\left(  x\left(  C\right)  ,z\left(  C\right)  \right)  $,
they are%
\[
C_{1}=\left(  \frac{F_{0}}{\sqrt{3}}\left(  2+\lambda\right)  ,0\right)  ~,
\]%
\[
C_{2}=\left(  x_{2},\sqrt{F_{0}-\left(  x_{2}\right)  ^{2}}\right)  ~.
\]

Point $C_{1}$ describes a universe in which the field $\phi$ and the potential
function contribute to the cosmological solution. The physical parameter
$w_{eff}$ is calculated to be $w_{eff}\left(  C_{1}\right)  =\frac{1}%
{3}\left(  2F_{0}\left(  \lambda^{2}-4\right)  -3\right)  $ from which it
easily follows that $C_{1}$ describes an accelerated asymptotic cosmological
solution for $\left\{  \left\vert \lambda\right\vert <2,F_{0}>\frac{1}%
{\lambda^{2}-4}\right\}  $ or $\left\{  \left\vert \lambda\right\vert
>2,F_{0}<\frac{1}{\lambda^{2}-4}\right\}  $ while, for $\left\vert
\lambda\right\vert =2$, the de Sitter Universe is recovered. The eigenvalues
of the linearized system near to the stationary point are%
\begin{equation}
e_{1}\left(  C_{1}\right)  =-3+F_{0}\left(  \lambda^{2}+2\right)
~,~e_{2}\left(  C_{1}\right)  =-3+F_{0}\left(  \lambda^{2}+2\right)  \text{~}.
\end{equation}
Thus $C_{1}$ is an attractor when $-3+F_{0}\left(  \lambda^{2}+2\right)  <0$.

\begin{figure}[ptb]
\centering\includegraphics[width=1\textwidth]{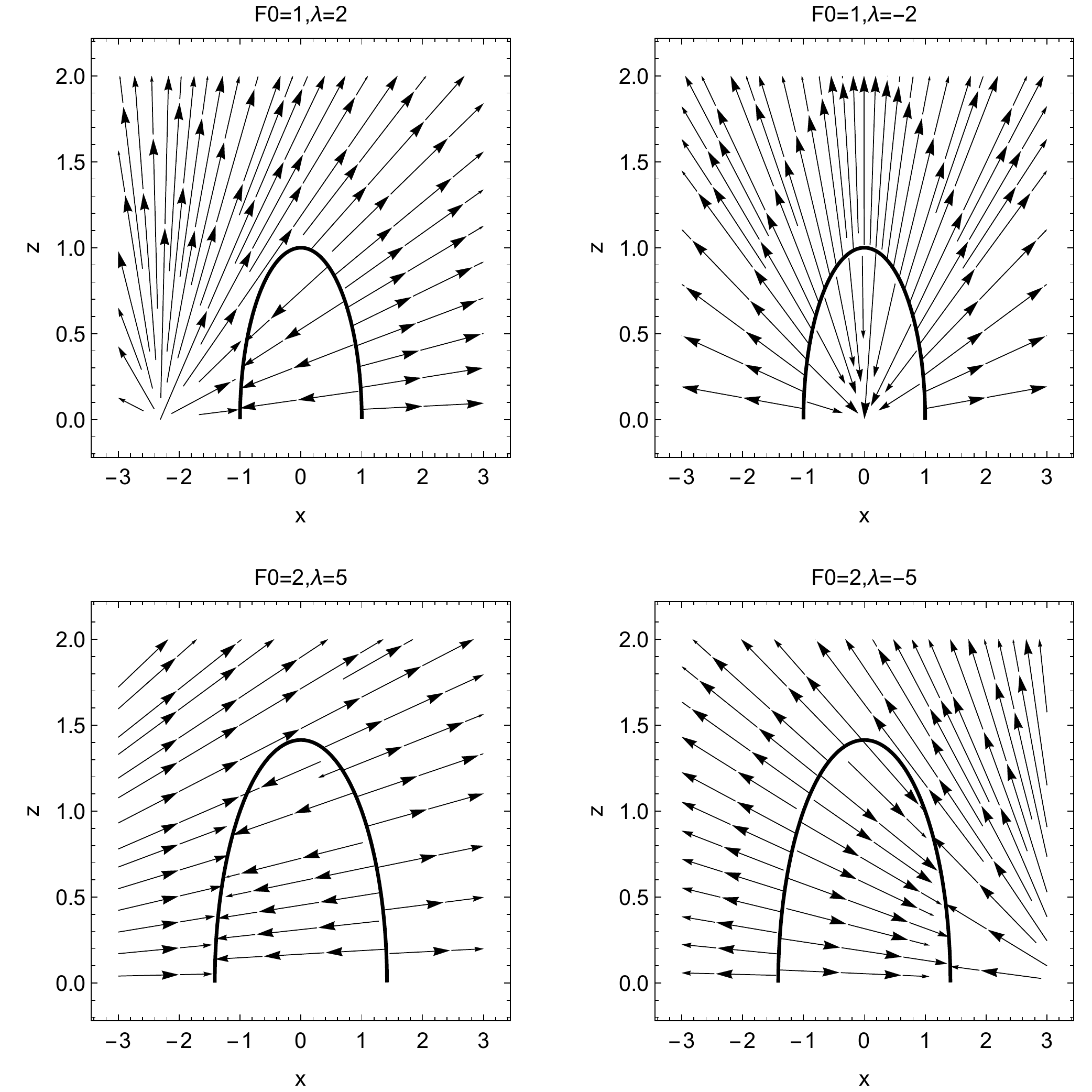}\caption{Phase-space
portrait for the scalar-torsion theory on the two-dimensional plane $\left(
x,z\right)  ,$ for different values of the free parameters $F_{0}$ and
$\lambda$ in order to investigate the stability properties of the family of
points $C_{2}$ (solid lines). From the plots we observe that, when
$e_{1}\left(  C_{2}\right)  <0$, the points ae attractors, otherwise they are
source points}%
\label{ff21}%
\end{figure}

The family of points $C_{2}$ exists when$~F_{0}>0$ and describes asymptotic
solutions with $w_{eff}\left(  C_{2}\right)  =1+\frac{8}{\sqrt{3}}x_{2}$ and
eigenvalues%
\[
e_{1}\left(  C_{2}\right)  =6+2\sqrt{3}\left(  2+\lambda\right)  x_{2}%
~,~e_{2}\left(  C_{2}\right)  =0.
\]
Because one of the eigenvalues is zero, in Fig. \ref{ff21} we present the
phase-space portraits for the dynamical system in the two-dimensional plane
$\left(  x,z\right)  $ for positive values of $F_{0}$ and different parameters
of $\lambda$. From the diagrams we observe that when $e_{1}\left(
C_{2}\right)  <0$ the family of points $C_{2}$ describes attractors, otherwise
the points are sources.

\subsection{Analysis at Infinity}

We consider now the Poincar\'e map (\ref{ai.P}) and we write the
two-dimensional system in the equivalent form%
\begin{equation}
F_{0}\frac{dX}{d\sigma}=\left(  F_{0}-\left(  1+F_{0}\right)  \left(
X^{2}+Z^{2}\right)  \right)  \left(  \sqrt{3}F_{0}\left(  2+\lambda\right)
\left(  X^{2}-1\right)  -3X\sqrt{1-X^{2}-Z^{2}}\right)  ~,
\end{equation}%
\begin{equation}
F_{0}\frac{dZ}{d\sigma}=\left(  F_{0}-\left(  1+F_{0}\right)  \left(
X^{2}+Z^{2}\right)  \right)  \left(  \sqrt{3}F_{0}\left(  2+\lambda\right)
X-3\sqrt{1-X^{2}-Z^{2}}\right)  .
\end{equation}

The stationary points at infinity are $D=\left(  X\left(  D\right)  ,Z\left(
D\right)  \right)  $ with $1-X\left(  D\right)  ^{2}-Z\left(  D\right)
^{2}=0$. We find the points
\begin{equation}
D_{1}=\left(  1,0\right)  \text{ and }D_{2}=\left(  -1,0\right)  \text{,}%
\end{equation}
from which it is clear that only the scalar field $\phi$ contributes in the
cosmological solution.

The eigenvalues of the linearized system are%
\begin{equation}
e_{1}\left(  D_{1}\right)  =-\sqrt{3}\left(  2+\lambda\right)
~,~\operatorname{Re}\left(  e_{2}\left(  D_{1}\right)  \right)  \simeq
-sign\left(  F_{0}\right)
\end{equation}
and%
\begin{equation}
e_{1}\left(  D_{2}\right)  =\sqrt{3}\left(  2+\lambda\right)
~,~\operatorname{Re}\left(  e_{2}\left(  D_{2}\right)  \right)  \simeq
-sign\left(  F_{0}\right)  .
\end{equation}

Therefore point $D_{1}$ is an attractor when $F_{0}>0$ and $2+\lambda>0$,
while $D_{2}$ is attractor for $F_{0}>0$ and $2+\lambda<0$.

\begin{figure}[ptb]
\centering\includegraphics[width=1\textwidth]{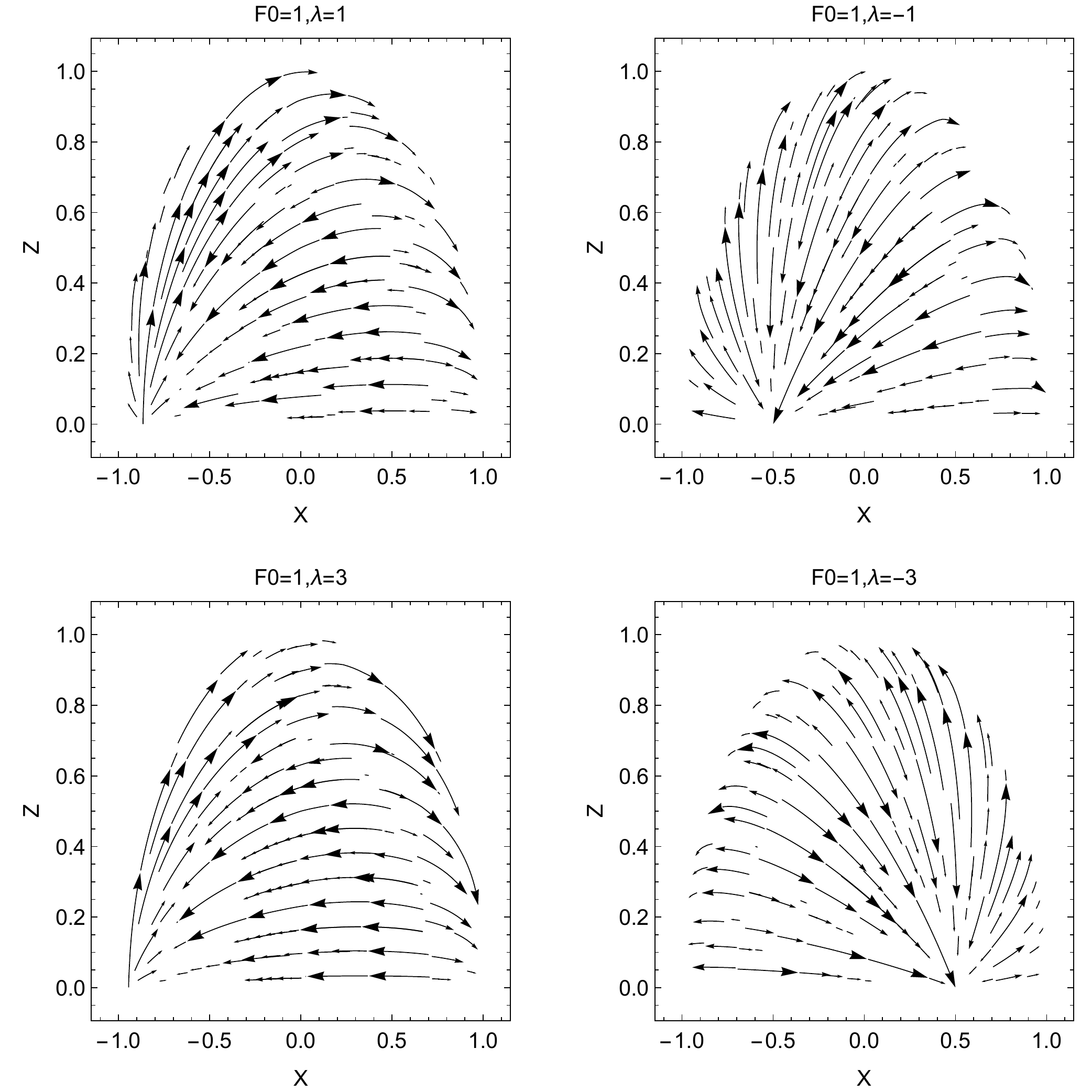}\caption{Phase-space
portraits for the scalar-torsion theory on the two-dimensional plane $\left(
X,Z\right)  ,$ for different values of the free parameters $F_{0}$ and
$\lambda$.}%
\label{ff22}%
\end{figure}

In Fig. \ref{ff22} we present phase-space portraits for the dynamical system
of scalar-torsion theory on the Poincar\'e variables.

\section{Conclusions}

\label{conc}

In this piece of work we considered a complex scalar field in the context of
scalar-tensor and scalar-torsion theories in a spatially flat FLRW background
geometry. For these two gravitational models we derived the field equations
and we investigated the dynamical evolution of the physical quantities by
using dimensional variables. With the use of the latter variables the field
equations for both theories are reduced to a two-dimensional dynamical system
of first-order ordinary differential equations. In order to construct the
cosmological history as provided by the given models we determined the
stationary points and we investigated their stability properties.

For the two different dynamical systems which correspond to the scalar-tensor
and scalar-torsion theories respectively, we found that the stationary points
for the two dynamical systems are four. They are two points at the finite
regime and two points at infinity. The stationary points can describe
accelerated asymptotic solutions which can describe the early and late-time
acceleration phases of the universe. The stationary points and their stability
properties are summarized in Table \ref{tab1}.

We can easily conclude that the two theories, that is, the scalar-torsion
theory and the scalar-tensor theory have the same number of stationary points
and similar stability properties. The physical variables of the background
geometry at the stationary points have similar properties with a different
functional dependence on the free variables $F_{0}$ and $\lambda$. The
dynamical equivalence of the two theories is an expected result. Torsion is
not dynamical and there are no new degrees of freedom in scalar-torsion
cosmology with respect to scalar-tensor cosmology.

From the above analysis it is clear that the two theories cannot be
distinguished from the evolution of the background geometry and the analysis
of the perturbations should be performed. However, that overpasses the scopus
of this work and will be published elsewhere.

Finally, as far as the complex scalar field is concerned, we remark that there
only at the (families) stationary points $A_{2}$ and $C_{2}$ the imaginary
part of the complex scalar field contributes in the physical quantities of the
asymptotic solutions. However, these families of points can describe important
eras of the cosmological history, for instance, the early inflationary era,
the radiation era or the matter dominated era. What will be of special
interest is to investigate the existence of a scalar field potential which
does not depend only on the norm of the complex scalar field.%

\begin{table}[tbp] \centering
\caption{Stationary points and their stability properties for the cosmological models of our consideration.}%
\begin{tabular}
[c]{cccc}\hline\hline
\textbf{Theory} & \textbf{Point} & \textbf{Finite/Infinity} &
\textbf{Stability}\\\hline
\textbf{Scalar-Tensor} &  &  & \\
& $A_{1}$ & Finite & Attractor see Fig. \ref{ff01}\\
& $A_{2}$ & Finite & Saddle\\
& $B_{1}$ & Infinity & Attractor $\frac{\sqrt{3}\left(  \lambda-4\right)
}{12F_{0}-1}<0$\\
& $B_{2}$ & Infinity & Attractor $-\frac{\sqrt{3}\left(  \lambda-4\right)
}{12F_{0}-1}<0$\\
&  &  & \\
\textbf{Scalar-Torsion} &  &  & \\
& $C_{1}$ & Finite & Attractor $-3+F_{0}\left(  \lambda^{2}+2\right)  <0$\\
& $C_{2}$ & Finite & Attractor $6+2\sqrt{3}\left(  2+\lambda\right)  x_{2}%
<0$\\
& $D_{1}$ & Infinity & Attractor $F_{0}>0$, $2+\lambda>0$\\
& $D_{2}$ & Infinity & Attractor $F_{0}>0$, $2+\lambda>0$\\\hline\hline
\end{tabular}
\label{tab1}%
\end{table}%

\bigskip

\acknowledgements

This work was partially supported by the National Research Foundation of South
Africa (Grant Numbers 131604). The author thanks Dr. G.\ Anargirou for the
hospitality provided while part of this work carried out.


\bigskip

\begin{thebibliography}{99}                                                                                               %


\bibitem {ratra}B. Ratra and P.J.E Peebles, Phys. Rev. D 37 3406 (1988)

\bibitem {in0}J. Rubio,\ Front. Astron. Space Sci. 5, 50 (2019)

\bibitem {in1}L.N. Granda and D.F. Jimenez, EPJC 79, 772 (2019)

\bibitem {in2}P. Parsons and J.D. Barrow, Phys.\ Rev. D 51, 6757 (1995)

\bibitem {in3}J.D. Barrow and A. Paliathanasis, Gen.\ Rel. Grav. 50, 82 (2018)

\bibitem {sf1}E.V Linder, Phys. Rev. D. 70 023511 (2004)

\bibitem {sf2}S. Barshay and G. Kreyerhoff, EPJC 5, 369 (1998)

\bibitem {sf3}T. Matos, J.-R. Luevano, I. Quiros, L.A. Urena-Lopez and J.A.
Vazquez, Phys. Rev.\ D 80, 123521 (2009)

\bibitem {sf4}A.R. Liddle, C. Pahud and L.A. Urena-Lopez, Phys.\ Rev. D 77,
121301 (2008)

\bibitem {sf5}L.P. Chimento, V. Mendez and N. Zuccala, Class. Quantum Grav.
16, 3749 (1999)

\bibitem {sf6}C. Deffayet, S. Deser and G. Esposito-Farese, Phys. Rev. D 80,
064015 (2009)

\bibitem {sf7}D. Bertacca, N. Bartolo and S. Matarrese, Adv. Astron. 2010,
904379 (2010)

\bibitem {sf8}D. Bertacca, A. Raccanelli, O.F. Piatella, D. Pietrobon, N.
Bartolo, S. Matarrese and T. Giannantonio, JCAP 03, 039 (2011)

\bibitem {sf9}A. Su\'{a}rez and T. Matos, MNRAS 416, 87 (2011)

\bibitem {sf10}C. Cosme, J.G.\ Rosa and O. Bertolami, Phys. Lett. B 781, 639 (2018)

\bibitem {qq1}W. Hu, Phys. Rev. D 71, 047301 (2005)

\bibitem {hy4}D. Wands, Lect. Notes Phys. 738, 275 (2008)

\bibitem {cc1}A.A. Coley and R.J. van den Hoogen, Phys.\ Rev. D 62, 023517
(2000)\qquad

\bibitem {mm2}A. Arbey, Phys. Rev. D 74, 043516 (2006)

\bibitem {mm3}C.-J. Gao and Y.-G. Shen, Phys. Lett. B 541, 1 (2002)

\bibitem {mm4}D.S.M. Alves and G.M. Kremer, JCAP 10, 009 (2004)

\bibitem {mm5}V. Sivanesan, Phys. Rev. D 90, 104006 (2014)

\bibitem {mm6}A. Paliathanasis, Universe 8, 325 (2022)

\bibitem {com1}I.M. Khalatnikov and A. Mezhlumian, Phys. Lett. A 169, 308 (1992)

\bibitem {com2}I.M. Khalatnikov, Lect. Notes Phys. 455, 343 (1995)

\bibitem {com3}D. Scialom, Helv. Phys. Acta 69, 190 (1996)

\bibitem {com4}P. Jetzer and D. Scialom, Phys.\ Rev. D 55, 7440 (1997)

\bibitem {com5}J.-A. Gu and W.-Y. Hwang, Phys. Lett. B 517, 1 (2001)

\bibitem {com6}A.Yu.Kamenshchik, I.M.Khalatnikov and A.V.Toporensky, Phys.
Lett. B 357, 36 (1995)

\bibitem {com7}M. Arik, M. Calik and N. Katirci, Central Eur. J. Phys. 9, 1465 (2011)

\bibitem {com8}G. Rosen, EPL 89, 19002 (2010)

\bibitem {com9}Y.-G. Shen and X.-H. Ge, Int. J.\ Theor. Phys. 45, 17 (2006)

\bibitem {com10}A. Arbey, Phys. Rev. D 74, 043516 (2006)

\bibitem {com11}R.C.G. Landim, EPJC 76, 1 (2016)

\bibitem {com12}B. Li, P.R. Shapiro and T. Rindler-Daller, PoS BASH2015, 028 (2016)

\bibitem {com13}A. Bevilacqua, J. Kowalski-Glikman and W. Wislicki,
Phys.\ Rev. D 105, 105004 (2022)

\bibitem {com14}H. Foidl and T. Rindler-Daller, \ Phys. Rev. D 105, 123534 (2022)

\bibitem {sc1}V. Faraoni, Cosmology in Scalar-Tensor Gravity, Fundamental
Theories of Physics, Springer Dordrecht (2004)

\bibitem {Brans}C. Brans and R.H. Dicke, Phys. Rev. 124, 195 (1961)

\bibitem {Jord}P. Jordan, Nature 164, 637 (1937) and \textit{Schwerkraft und
Weltfall,} 2nd ed., Vieweg und Sohn, Braunschweig, (1955)

\bibitem {Hayashi79}K. Hayashi and T. Shirafuji, Phys. Rev. D 19, 3524 (1979)

\bibitem {Tsamp}M. Tsamparlis, Phys. Lett. A 75, 27 (1979)

\bibitem {omegaBDGR}V. Faraoni, Phys. Rev. D 59, 084021, (1999)

\bibitem {st1}J. Garcia-Bellido, D. Wands, Phys. Rev. D 52, 6739 (1995)

\bibitem {st2}S.Sen and A.A. Sen, Late time acceleration in Brans Dicke
Cosmology, Phys.\ Rev. D 63, 124006 (2001)

\bibitem {st3}M. Artymowski, Z. Lalak and M. Lewicki, Inflation and dark
energy from the Brans-Dicke theory, JCAP 06, 031 (2015)

\bibitem {st4}M. Gasperini and G. Veneziano, Astropart. Phys. 1, 317 (1993)

\bibitem {st5}L. Jarv and J. Lember, Universe 7, 178 (2021)

\bibitem {st6}M. Hohmann, Phys.\ Rev. D 98, 064002 (2018)

\bibitem {st7}M. Hohmann and C. Pfeifer, Phys.\ Rev. D 98, 064003 (2018)

\bibitem {st8}H. Wei, Phys. Dynamics of Teleparallel Dark Energy, Lett. B 712,
430 (2012)

\bibitem {te2}C.-Q.\ Geng, C.-C. Lee, E.N. Saridakis and Y.-P. Wu,
"Teleparallel" Dark Energy, Phys. Lett. B 704, 384 (2011)

\bibitem {te3}C. Xu, E.N. Saridakis and G. Leon, Phase-Space analysis of
Teleparallel Dark Energy, JCAP 07, 005 (2012)

\bibitem {te4}A. Paliathanasis, Eur. Phys. J. Plus 136, 674 (2021)

\bibitem {dyn1}E.J. Copeland, A.R. Liddle and D.\ Wands, Phys. Rev. D
\textbf{57}, 4686 (1998)

\bibitem {dyn2}. Wainwright and G.F.R Ellis, Cambridge University Press,
Cambridge (1997)

\bibitem {dyn3}A.A. Coley, Dynamical Systems and Cosmology, Springer,
Dordrecht (2003)

\bibitem {dyn4}L. Amendola and S. Tsujikawa, Dark Energy, Cambridge University
Press, Cambridge (2010)

\bibitem {dyn6}C.R. Fadragas and G. Leon, Class. Quantum Grav. 31, 195011 (2014)

\bibitem {dyn5}O. Hrycyna and M. Szydlowski, JCAP 1312, 016 (2013)

\bibitem {revtel}S. Bahamonte, K.F. Dialektopoulos, C. Escamilla-Rivera, V.
Gakis, M. Hendry, J.L. Said, J. Mifsud and E. Di Valentino, Teleparallel
Gravity: From Theory to Cosmology, [arXiv:2106.13793] (2021)
\end{thebibliography}
\end{document}